\documentclass[11p,a4paper]{article} 
\pdfoutput=1
\usepackage{jinstpub}
\usepackage{mathrsfs}
\usepackage{siunitx}
\usepackage{subcaption}

\title{Field dependence of electronic recoil signals in a dual-phase liquid xenon time projection chamber}
\author[1]{E. Hogenbirk,\note{Corresponding author.}}
\author{M. P. Decowski,} 
\author{K. McEwan,} 
\author{A. P. Colijn} 

\affiliation{Nikhef and the University of Amsterdam, Science Park, 1098XG Amsterdam, Netherlands}
\emailAdd{ehogenbi@nikhef.nl}

\arxivnumber{1807.07121}

\abstract{
We present measurements of light and charge signals in a dual-phase time projection chamber at electric fields varying from \SI{10}{V/cm} up to \SI{500}{V/cm} and at zero field using \SI{511}{keV} gamma rays from a $^{22}$Na source.
We determine the drift velocity, electron lifetime, diffusion constant, and light and charge yields at \SI{511}{keV} as a function of the electric field.
In addition, we fit the scintillation pulse shape to an effective exponential model, showing a decay time of \SI{43.5}{ns} at low field that decreases to \SI{25}{ns} at high fields.
}


\notoc 

\begin{document}
\maketitle
\section{Introduction}\label{sec:introduction}
The search for weakly interacting massive particles by direct detection has seen sensitivity improvements of orders of magnitude in recent years, chiefly due to the employment of dual-phase liquid xenon time projection chambers (TPCs). 
In this type of detector, two signals following an energy deposition in the liquid xenon are registered, the first due to the scintillation light, called S1, and the second due to the ionized electrons, called S2.
For the S2, the electrons liberated in the liquid xenon drift up towards a gas layer under the influence of an electric field, where a stronger electric field extracts them from the liquid and produces proportional scintillation.
The light from both signals is detected using photomultiplier tubes (PMTs).

The magnitude of the applied drift field influences the performance of TPCs, as the dynamical behavior of the free electrons is changed.
Firstly, for lower fields, the electrons are more likely to recombine with xenon ions, forming excitons that decay and contribute to the scintillation signal instead of the ionization signal.
In effect, this changes the ratio of S2/S1 as well as the scintillation pulse shape due to the time delay in exciton formation.
Secondly, the drift field determines the electron drift velocity, which rises steeply at fields up to  $\sim$\SI{100}{V/cm} and saturates for higher fields.
Thirdly, longitudinal electron diffusion depends on the applied field, so that S2 signals become wider at lower fields for the combination of two reasons: a lower drift velocity (and thus more time for electron diffusion for a given drift length) and a higher diffusion constant.
Finally, the electron lifetime, the average time before an electron is absorbed in the liquid xenon, is assumed to depend on the electric field \cite{aprile2006b, aprile2010}, although direct measurements of this are scarce \cite{aprile1991, ferella2006, ferella2007}.

As the size of liquid xenon TPCs has increased, it has become apparent that the applied drift fields have become progressively lower.
Longer drift lengths require higher cathode voltages, providing an engineering challenge.
For example, the XENON10, XENON100, LUX and XENON1T TPCs have operated at maximum average fields of \SI{730}{V/cm} \cite{angle2008, aprile2011}, \SI{530}{V/cm} \cite{aprile2012}, \SI{181}{V/cm} \cite{akerib2014} and \SI{117}{V/cm} \cite{aprile2017}, respectively. 
Planned future TPCs, such as XENONnT \cite{aprile2016}, LZ \cite{LZ} and DARWIN \cite{aalbers2016} will feature even longer drift lengths, requiring vastly improved high voltage engineering or operating at fields lower than \SI{100}{V/cm}.
Some of the aforementioned low drift field effects, such as the light and charge yields, are well-measured and described by the simulation toolkit NEST \cite{nest2011, nest2}, while other effects have not been systematically measured yet.

This work describes measurements with XAMS \cite{hogenbirk2016}, a dual-phase liquid xenon TPC, operating at fields between approximately \SI{10}{V/cm} and up to \SI{500}{V/cm}, as well as measurements at zero field.
We use data from \SI{511}{keV} gamma-ray recoils from a $^{22}$Na positron annihilation source.
The drift velocity, electron lifetime, diffusion constant, and light and charge yields are determined for all measured fields and are compared to NEST where possible.
In addition, we fit the scintillation pulse shape using a model with two exponential decay components \cite{hogenbirk2018}, where we allow the effective triplet lifetime component to vary.

\section{Measurements}
\subsection{Data acquisition and processing}
The active volume of the XAMS TPC has a cylindrical geometry, with the distance of the gate to cathode mesh of \SI[separate-uncertainty=true]{100 \pm 2}{mm} and a diameter of \SI{44}{mm} (at \SI{-90}{\celsius}).
S1 and S2 photons are detected with two 2-inch R6041-406 Hamamatsu PMTs \cite{hamamatsu} that view the active volume from above and below.
The high voltage required for the PMT bias, anode and cathode are provided by a CAEN DT1471ET power supply.
During all measurements, the PMT voltages were \SI{750}{V} and \SI{700}{V} for the bottom and top PMT, respectively, resulting in a gain of \num{2.69e5} and \num{2.90e5} for the bottom and top PMT, respectively.
This relatively low value was set to avoid signal saturation.
The extraction field was supplied by biasing the anode with \SI{3.5}{kV} for all measurements, except for the zero field measurement.

The gamma-ray source is a \SI[separate-uncertainty=true]{172 \pm 5}{kBq} $^{22}$Na source, that decays by positron emission and subsequent \SI{1274}{keV} gamma ray with a branching ratio of \SI{90.4}{\%}.
The positron rapidly annihilates and produces two back-to-back \SI{511}{keV} gamma rays.
We use an external NaI(Tl) detector to trigger on one of the three gamma rays, with one of the other gamma rays interacting in the TPC.
If a three-fold coincidence of both PMTs in the TPC and the NaI(Tl)-detector is found within \SI{120}{ns}, full waveform signals are recorded for all channels.
We changed the waveform duration from \SI{164}{\micro s} (at high fields) up to \SI{328}{\micro s} to account for the reduced drift velocity at low fields.
For all except the lowest field measured, the event window was long enough to capture the entire drift length, with the S1 in the middle of the waveform.
The source was mounted in a lead collimator setup (similar as described in \cite{hogenbirk2016}) aimed at the bottom part of the TPC.
The $z$-position distribution of the interactions follows a Gaussian distribution at \SI[separate-uncertainty=true]{-58.6 \pm 16.6}{mm} ($\mu \pm \sigma$) (see section~\ref{sec:drift}). 
For the measurements described here, we varied the electric field strength by applying different voltages on the cathode.
A total of \num{10} voltage settings were used with voltages ranging from \SI{100}{V} to \SI{5000}{V}, giving fields up to \SI{500}{V/cm}.
In addition, we measured at zero applied voltage, where only the scintillation signal is produced.

\subsection{Event selection}
Valid events are selected using three cuts.
Only events with precisely one S1 and S2 are selected, removing a large fraction of events containing pileup signals and multiple scatter events. 
Events with high S2 widths are cut, as they are indicative of merged S2 signals coming from unresolved multiple scatters.
All events are further required to contain an S1-signal coincident with a signal in the NaI(Tl) detector.
The final fraction of events passing all cuts is \SI[separate-uncertainty=true]{57 \pm 2}{\%} for all measurements except for the lowest field, where the event window was not long enough to capture events from the bottom of the TPC resulting in only \SI{40}{\%} of the events passing cuts.
For some of the analyses described below, only the events with a $z$-coordinate within the collimated beam (\SI[separate-uncertainty=true]{-58.6 \pm 16.6}{mm}) are selected.
This cut accepts \SI[separate-uncertainty=true]{64 \pm 1}{\%} of all events passing previous cuts.
For the lowest field measurement, the selection is limited to $z \geq \SI{-60}{mm}$, giving an acceptance of \SI{49}{\%}.

\begin{figure}[h!]
\centering
\begin{subfigure}{.5\textwidth}
  \centering
  \includegraphics[width=\linewidth]{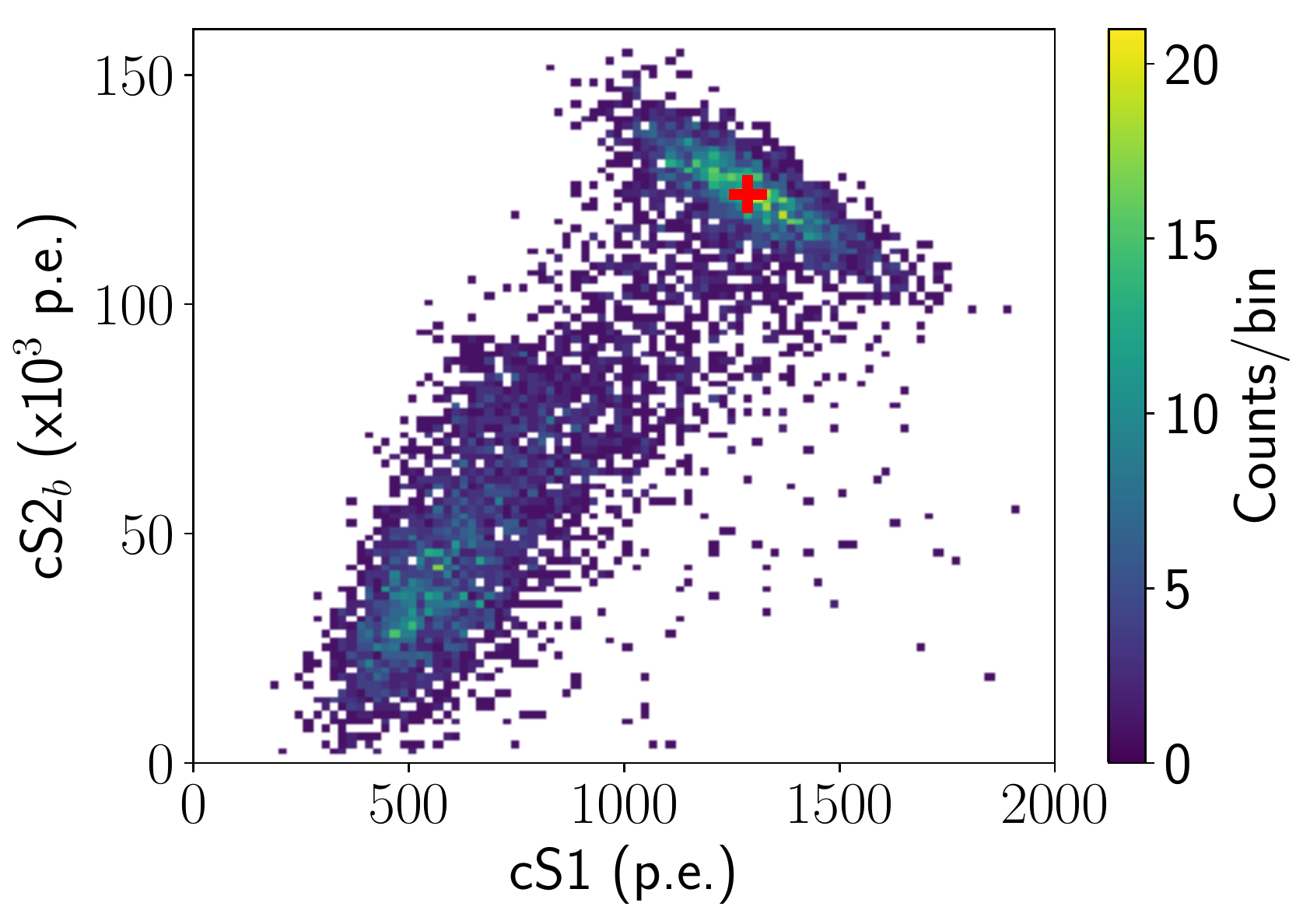}
  \caption{}
  \label{fig:cs1cs2}
\end{subfigure}%
\begin{subfigure}{.5\textwidth}
  \centering
  \includegraphics[width=\linewidth]{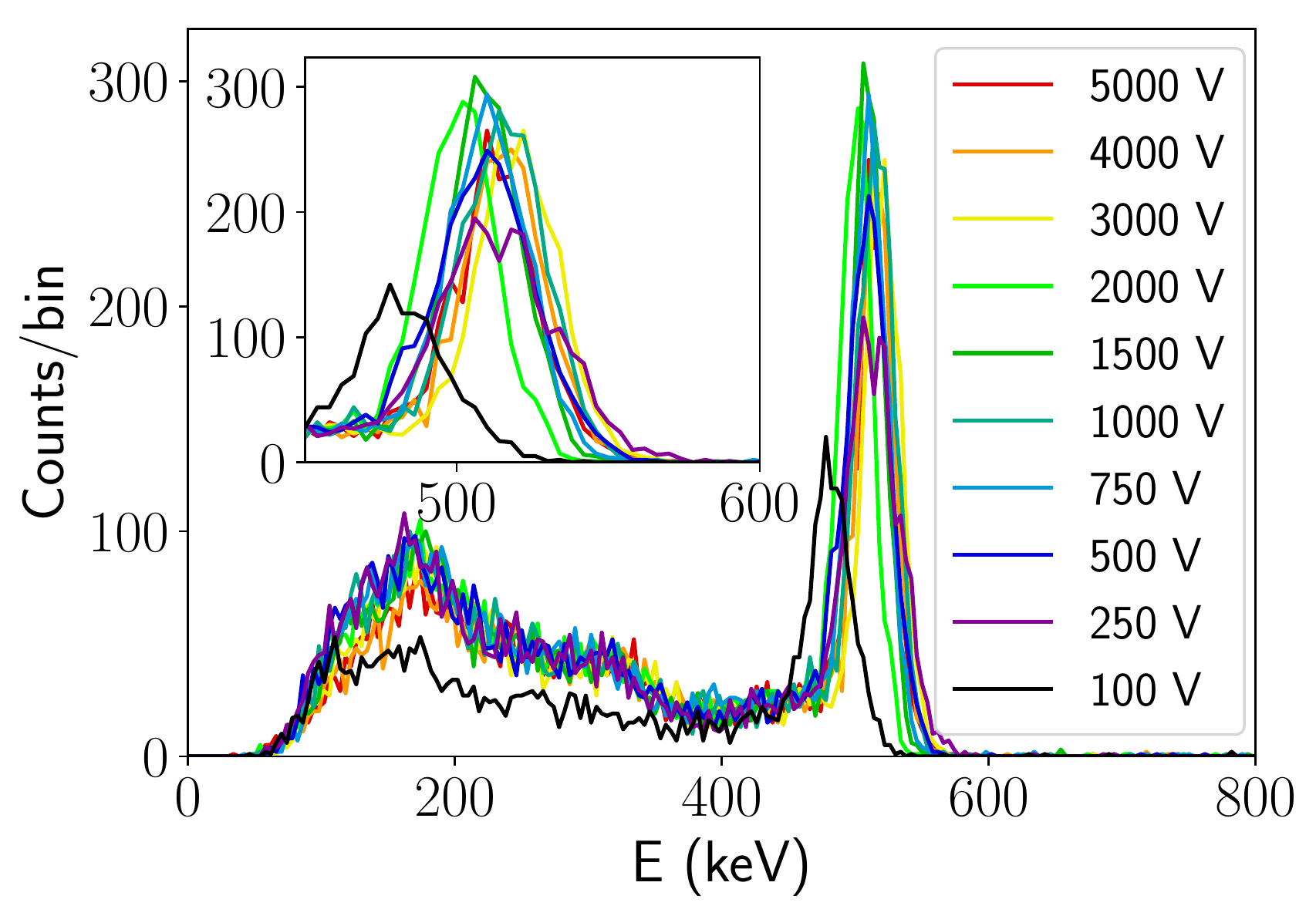}
  \caption{}
  \label{fig:energy}
\end{subfigure}
\caption{(a): The distribution of the S1 and S2 signals after all cuts and corrections for light detection efficiency (for S1) and electron lifetime (for S2) for the described $^{22}$Na source.
The red cross shows the photopeak position as determined from Gaussian fits (see section~\ref{sec:g1g2}).
The population at lower values of cS1 and cS2$_b$ is due to Compton scatter events, which give a smaller energy deposition.
No events are observed at very low values of S1 due to the trigger threshold.
(b): The energy spectrum, reconstructed using a linear combination of S1 and S2, for all data.
The inset shows a detailed view of the photopeak at \SI{511}{keV}.
}
\label{fig:test}
\end{figure}

An example of the energy distribution in corrected S1 (cS1) and the corrected S2 using only the bottom PMT (cS2$_b$) is shown in figure~\ref{fig:cs1cs2} for a cathode voltage of~\SI{5000}{V}.
This distribution is plotted after applying all the aforementioned cuts and after correcting for the $z$-dependent light detection efficiency for S1 and for the electron lifetime in S2 (see section~\ref{sec:electron_lifetime}).
The \SI{511}{keV} photopeak is found at the position of the red cross, showing a downwards sloping ellipse due to the anti-correlation of S1 and S2.
An additional distribution due to Compton scatter events extends down to lower energies.
At low S1 values, no more events are found due to the trigger cutoff for low amplitude S1s.
For lower applied cathode voltages, the S1 increases in favor of the S2 signal, yielding a similar plot as figure~\ref{fig:cs1cs2} but shifted down in S2 and up in S1.

Since the S1 and S2 signals are anti-correlated, a superior energy resolution can be achieved with a linear combination of the two signals.
This is called the combined energy scale (see section~\ref{sec:g1g2}).
In figure~\ref{fig:energy}, the combined energy spectrum is shown for all data.
All measurements except for the one taken at \SI{100}{V} show a very similar energy spectrum (the discrepancy of the \SI{100}{V} datapoint is discussed in section~\ref{sec:g1g2}) with a clear photopeak at \SI{511}{keV}.
The energy resolution achieved at this energy is \SI[separate-uncertainty=true]{2.8 \pm 0.5}{\%}, where the uncertainty indicates the standard deviation across different measurements.

\subsection{Electric field simulation}
\begin{figure}[h]
\begin{center}
\includegraphics[width=0.7\linewidth]{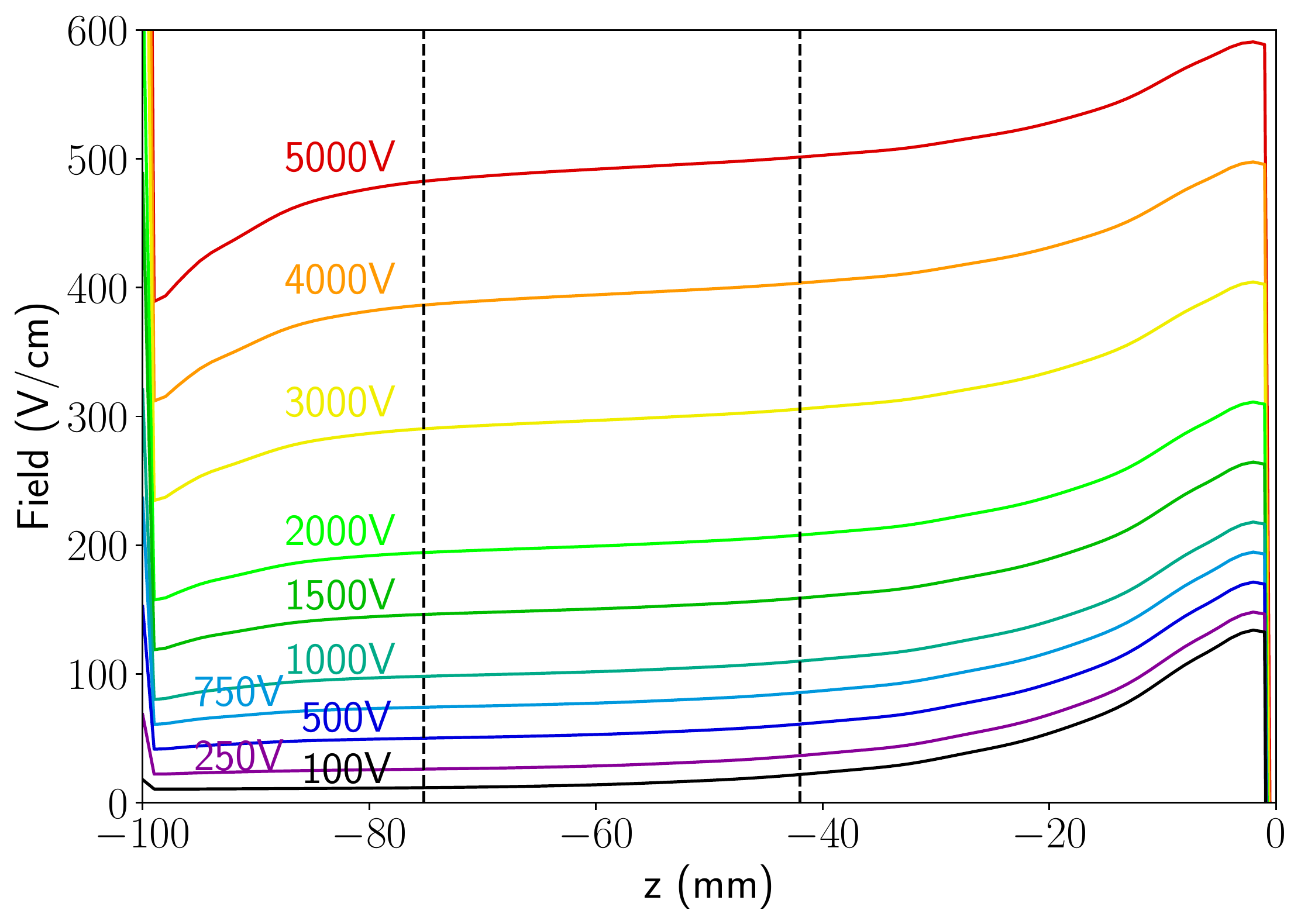}
\caption{Electric field as a function of the $z$-coordinate for various applied cathode voltages, calculated using Comsol Multiphysics \cite{comsol}.
The gate mesh is located at $z=0$, and the cathode is at \SI{-100}{mm}.
The dotted lines indicate the region containing the collimated beam ($\mu \pm 1\sigma$).
}
\label{fig:fields}
\end{center}
\end{figure}

The drift field in the TPC is calculated using the Comsol Multiphysics package \cite{comsol}.
The model, described in detail in \cite{hogenbirk2014}, uses the cylindrical symmetry of the TPC and includes the geometry and electrical properties of the Teflon structure, the meshes and the vessel holding the TPC.
Figure~\ref{fig:fields} shows the resulting electric field as a function of the $z$-coordinate for all used voltages.
This gives the field averaged over $r^2$ (volume average), as the dependence on $r$ is relatively minor.
There are two regions of high field distortion: at the top, caused by the high anode voltage `leaking' through the grounded gate mesh at the top of the TPC ($z = 0$), and at the bottom, caused by the high cathode voltage.
The region where the collimated beam is located, indicated by the dashed lines, is a region of relatively low field distortion.
We calculate the average field and its uncertainty by taking the volume-averaged field and its standard deviation in the relevant regions.
Very close to the meshes, the field model breaks down due to the assumed cylindrical symmetry which is incorrect for the square mesh structure.
We therefore restrict the average field computation to between \SI{-5}{mm} and \SI{-95}{mm}.

\section{Results}
\subsection{Drift velocity and $z$-coordinate} \label{sec:drift}
The drift velocity as a function of field is determined from the maximum observed drift time, which corresponds to the position of the cathode.
We subtract \SI[separate-uncertainty=true]{1.0 \pm 0.5}{\micro s} from this time to account for the finite drift time between the gate mesh and the liquid level, which we observe in the data by the changing S2/S1 ratio as a result of the high field between the gate mesh and anode mesh.
The uncertainty of \SI{0.5}{\micro s} on the drift time and the estimated uncertainty of \SI{2}{mm} of the distance from the cathode to gate are propagated to the drift velocity uncertainty.
We reconstruct the $z$-coordinate by assuming a constant drift velocity over the full drift region.
Although the calculated electric fields show non-uniformity in the drift region and therefore a non-uniform drift velocity, for high enough fields the drift velocity saturates so that the drift velocity is relatively constant over the full range.
The relative inhomogeneity in drift velocity in the range from \SI{-5}{mm} to \SI{-95}{mm} is \SI{1.5}{\%} for an applied voltage of \SI{5000}{V} and increases to \SI{7}{\%} for \SI{750}{V}, calculated assuming drift velocities from NEST.
Correcting for this effect would require a parameterization of $v_d(E)$ and an iterative approach, and was deemed unnecessary for this analysis.

For voltages of \SI{500}{V} and below, we observe a significant change in the $z$-positions reconstructed using the drift time with respect to the higher field measurements.
We attribute this to field inhomogeneity and the large gradient of the drift velocity as function of field below \SI{100}{V/cm}.
For these measurements, we reconstruct the $z$-coordinate by using the fraction of S1 light observed in the top PMT $f_t$.
We calibrate $f_t$ as a function of $z$ using the highest field measurement, where field distortion is minimal.
For the three lowest field measurements, we determine the drift velocity by taking the average drift times at two points with $f_t$-derived $z$-coordinates.
We estimate that using these coordinates introduces an uncertainty of \SI{2}{mm} in the drift distance and take this as the uncertainty on the drift velocity.
The resulting drift velocities are indicated in figure~\ref{fig:superplot}(a) and show good agreement with NEST.

\begin{figure}[h!]
\begin{center}
\includegraphics[width=0.88\linewidth]{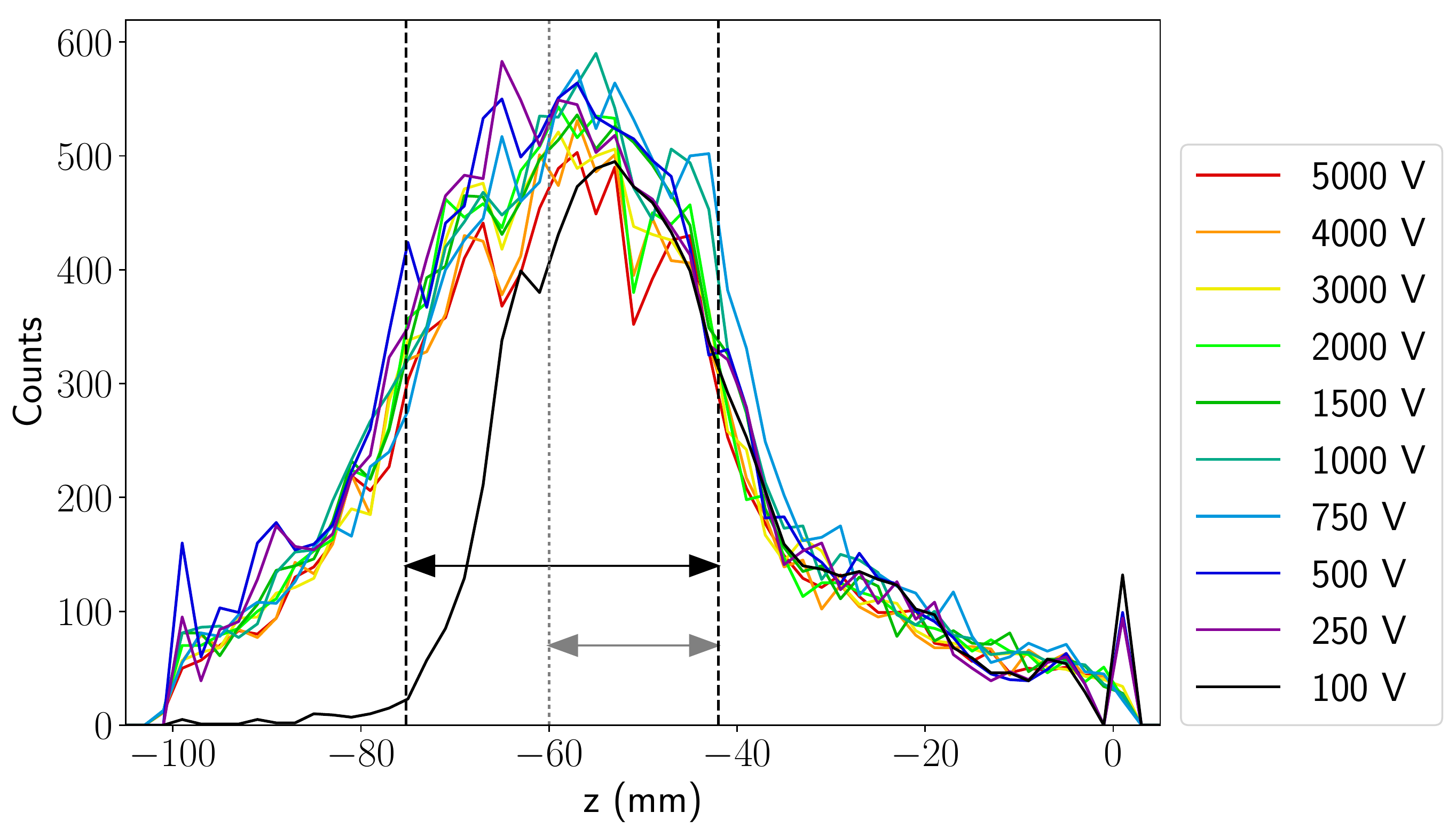}
\caption{
The $z$-distribution for all measurements, reconstructed using the drift time for measurements above \SI{500}{V} and using the S1 light distribution for $\SI{500}{V}$ and below.
The distributions show the same trend around the position of the collimated beam, except for the measurement at the lowest field due to an insufficiently long event window setting.
The black arrow indicates the selection range used in most of the analyses.
For the lowest field, the region indicated by the gray arrow is used instead. 
}
\label{fig:z_dist}
\end{center}
\end{figure}

The final position distribution for all measurements is shown in figure~\ref{fig:z_dist}.
The distributions show the same beam profile due to the collimator setup, except for the measurement at \SI{100}{V} that lacks events for low values of $z$ since the event window was too short to capture these events.
The dashed black lines indicate the boundaries of the position cut used in several of the following analyses.
For the \SI{100}{V} measurement, the dotted gray line is used as a lower bound.
The spiked feature at \num{0} and \SI{-100}{mm} for measurements at \SI{500}{V} and below are due to interpolation artifacts in the $f_t$-derived $z$-coordinate.

The S1-signal is corrected for $z$-dependent light detection efficiency~(LDE).
We use a previously determined LDE map (as described in~\cite{hogenbirk2018}) for this correction, where the corrected S1 is normalized to the volume-averaged S1.
We confirm that this LDE map is consistent with current conditions using a measurement where the $^{22}$Na source was uncollimated, so that the events are spread out over the full detector volume.

\subsection{Electron lifetime}\label{sec:electron_lifetime}
The attachment of free electrons to impurities in the liquid xenon causes a decrease of electrons with increasing drift time, which follows an exponential distribution characterized by the electron lifetime~$\tau$ according to
\begin{equation}
n_e(t_d) = n_{e, 0} \exp{\left(-\frac{t_d}{\tau}\right)}.
\end{equation}
We fit this function for events with a $z$-coordinate within \SI[separate-uncertainty=true]{58.6 \pm 16.6}{mm} of the collimated beam position.
To determine the electron lifetime, we select events within the full absorption peak, corresponding to a \SI{511}{keV} energy deposition, so that the mean initial number of electrons $\left< n_{e, 0} \right>$ is constant.
This is done iteratively.
Full absorption events are first selected using a Gaussian fit to the corrected S1 spectrum.
After S2 correction, the energy can be calculated using a combination of the S1 and S2 signals (see section~\ref{sec:g1g2}). 
The reconstructed energy is then used to refine the photopeak selection, after which the S2 correction is recomputed.
This process is repeated until the correction on the electron lifetime is small with respect to the fit uncertainty.

The electron lifetime as determined from a direct exponential fit is biased due to field inhomogeneity within the fit range.
Figure~\ref{fig:fields} shows that high $z$-coordinates correspond to a relatively higher field than events occurring deeper down in the TPC.
This implies that for high $z$-coordinates, more electrons are extracted from the interaction site, leading to a steeper observed fall of $n_e(t_d)$.
We calculate the magnitude of this effect by simulating the observed effective electron lifetime as a function of the true electron lifetime, assuming the field dependence of the S2 size as described by NEST.
We thus obtain a (higher) corrected electron lifetime and use this lifetime for the S2 correction.
In figure~\ref{fig:superplot}(b), the directly fitted and corrected electron lifetimes are shown with blue circles and green squares, respectively.
The indicated uncertainties correspond to the statistical uncertainty from the exponential fit.

The electron lifetime depends on the attachment cross-section of impurities in the liquid xenon, which depends on the electric field \cite{bakale1976, aprile2010}.
This cross-section may either increase or decrease with applied field, depending on the type of impurity.
For example, the lifetime increases with field for O$_2$ and H$_2$O, but decreases for N$_2$O and CO$_2$.
In this case, we find a maximum electron lifetime at fields between \num{100} and \SI{200}{V/cm}, decreasing at lower and higher fields.
However, it is unknown what impurity dominates absorption in our TPC, so that we are unable to extend this to a more general statement for other TPCs.

\subsection{Diffusion}

\begin{figure}[h!]
\begin{center}
\includegraphics[width=0.7\linewidth]{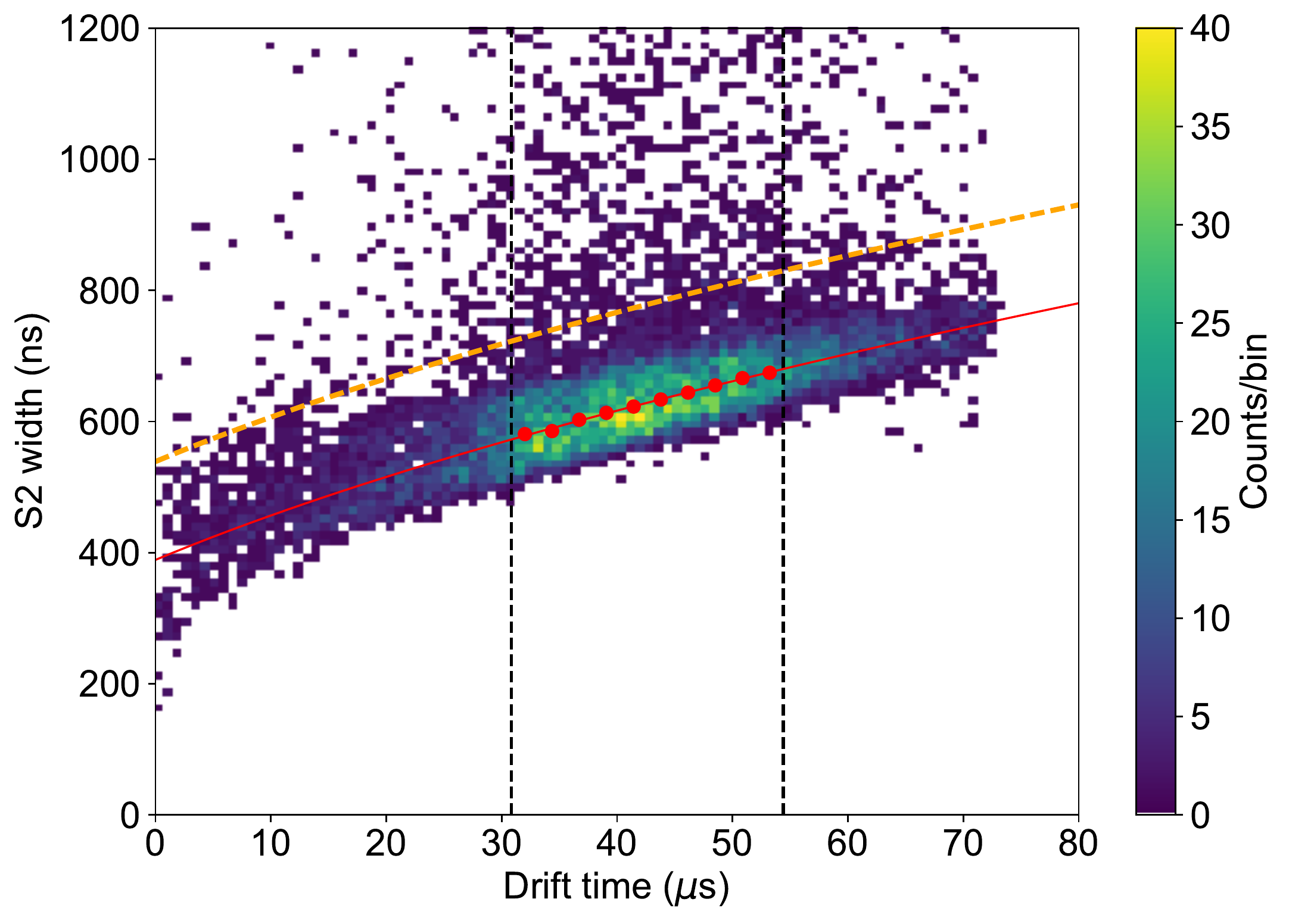}
\caption{
The S2 width as a function of drift time for the measurement taken at a cathode voltage of \SI{1000}{V}.
The increase in width is caused by diffusion and traces the square root fit that is shown by the red solid line.
This is fit to the median S2 width in drift time slices, as indicated by the red points.
}
\label{fig:diffusion}
\end{center}
\end{figure}

The width of an S2 increases for interactions occurring deeper in the TPC.
The standard deviation of the S2 time profile $\sigma_{\rm S2}$ can be described by
\begin{equation}
\sigma_{\rm S2} = \sqrt{\frac{2 D t_d(z)}{v_d^2} + \sigma_0^2} ,
\label{eqn_diff}
\end{equation}
with $D$ the diffusion constant, $t_d$ the drift time and $v_d$ the drift velocity.
The zero-drift S2 standard deviation~$\sigma_0$ comes from the S2 width due to the time electrons emit light within the gas gap.
For the measurements with applied voltages above \SI{500}{V}, we fit this function to the observed distribution of S2 width as a function of drift time.
An example of this fit (for \SI{1000}{V} cathode voltage) is shown in figure~\ref{fig:diffusion}.
The fit range, indicated by the black dotted lines, corresponds to the $z$-range shown in figure~\ref{fig:z_dist}.
We divide the range into \num{10} slices and compute the  median width in each slice.
This is indicated by the red points.
The solid red curve shows a fit of equation~\ref{eqn_diff} to the medians.
The dotted orange curve traces the fit plus a constant offset of \SI{150}{ns}; all events above this line are likely due to merged S2s from multiple scatter events and are cut in the analysis.

\begin{figure}[h!]
\begin{center}
\includegraphics[width=0.695\linewidth]{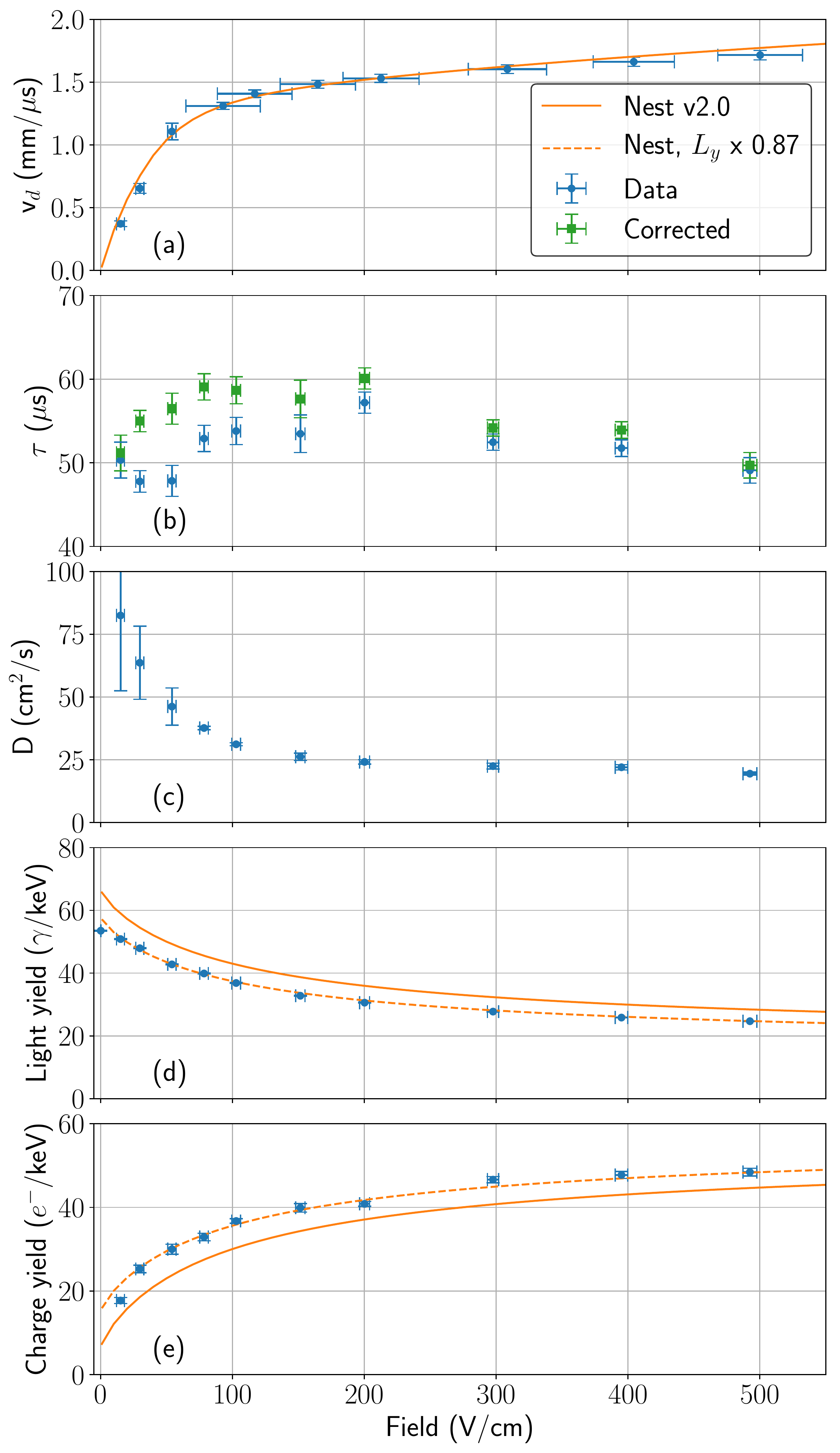}
\caption{
Various measured field-dependent properties compared to NEST predictions where available (orange lines) \cite{nest2011, nest2}. 
The panels show (a) the electron drift velocity, (b) electron lifetime, (c) diffusion constant, (d) light yield and (e) charge yield.
Blue points indicate the directly measured or fitted values, green points show values after corrections for field non-uniformity.
The dashed lines correspond to the NEST light yield decreased by \SI{13}{\%} and give a better fit to the measured light and charge yield.
The uncertainties in the field is evaluated as the standard deviation of the field within the $z$-selection.
For the drift velocity (a), the full $z$-range is used, giving a relatively large uncertainty.
}
\label{fig:superplot}
\end{center}
\end{figure}

For the three lowest fields, the aforementioned drift field inhomogeneity causes a significant deviation of this behavior due to the dependence of $D$ on the field, and thus on $z$. 
To account for this, we instead use
\begin{equation}
\frac{d\sigma_{\rm S2}^2}{dt_d} = \frac{2 D(z(E))}{v_d^2}.
\end{equation}
The left hand side is calculated by taking the difference between $\sigma_{\rm S2}^2$ in drift time slices, divided by their separation in drift time.
This has the advantage that the diffusion constant can be probed if it changes with on $z$, however, it is highly sensitive to uncertainties in $\sigma_{\rm S2}$.
This causes larger uncertainties of $D$ for the lowest field values.
We determine the uncertainty on $D$ calculated in this way from the standard deviation of $\Delta \sigma_{\rm S2}^2$ for several neighboring drift time slices, while the uncertainty of $D$ calculated with the direct fit of equation~\ref{eqn_diff} is the statistical uncertainty on the fit parameter.
The diffusion constant as a function of field is shown in figure~\ref{fig:superplot}(c).
The drift velocity used to compute $D$ is identical to the velocity shown in panel (a). 

For the measurements where we use the direct fit (above $\SI{500}{V}$), we estimate the impact of drift field inhomogeneity on the fitted value of $D$ by a simulation of the S2 width as a function of drift time.
This simulation uses the electric field as function of $z$ (see figure~\ref{fig:fields}), the field dependence of $v_d$ (from NEST) and an interpolation of $D$ as a function of field.
We then fit the simulation with equation~\ref{eqn_diff}, neglecting the effects of field inhomogeneity, and compare the result with the simulated value of $D$.
For the measurements above \SI{500}{V}, the difference is less than \SI{1}{\%}; well below the uncertainty on $D$.
We therefore conclude that the correction would be minor and that the field inhomogeneity can be neglected for these measurements.

\subsection{Light and charge yields}\label{sec:g1g2}
\begin{figure}[h]
\begin{center}
\includegraphics[width=0.7\linewidth]{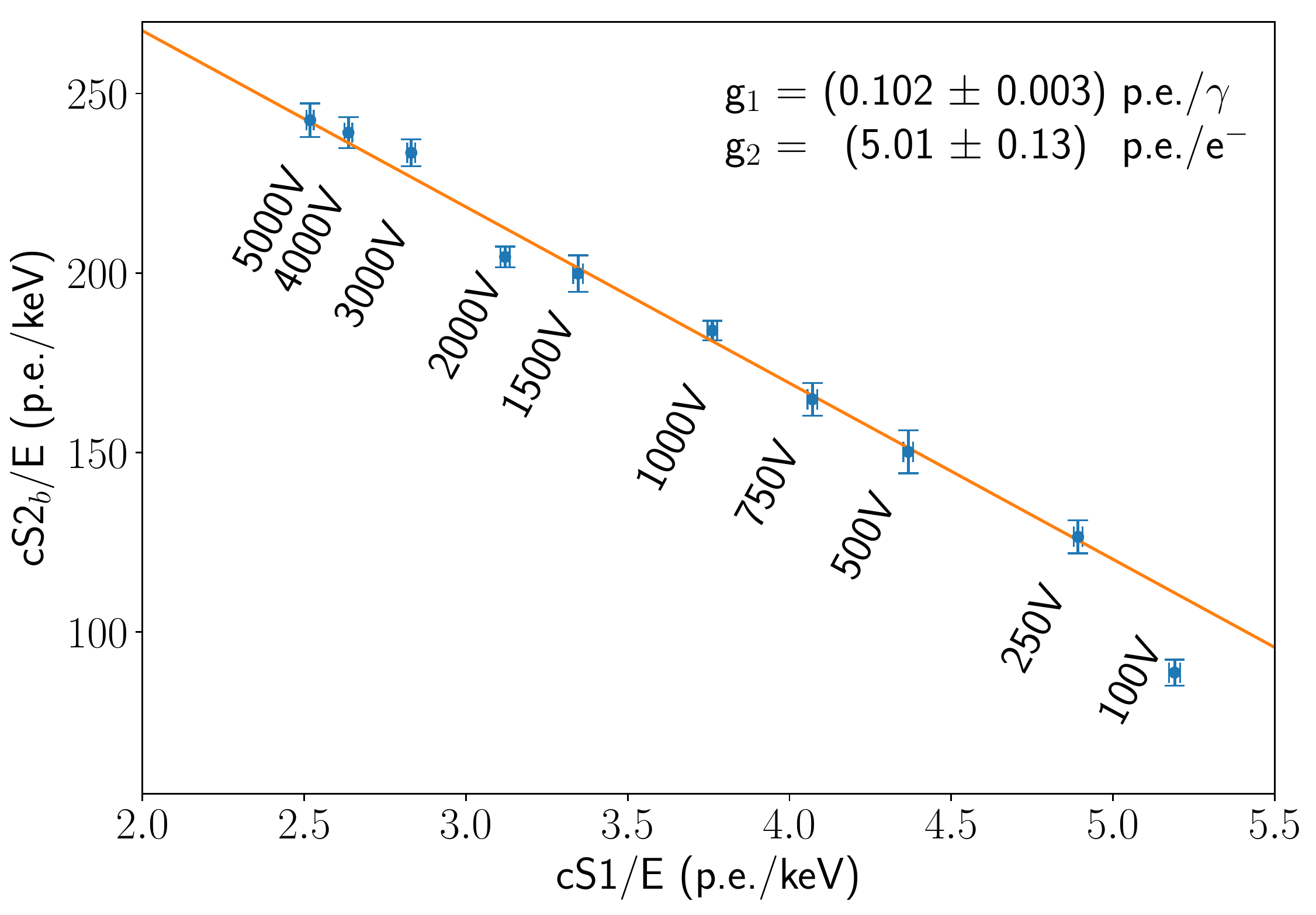}
\caption{The S2 and S1 yield, changing from high field (top left) to low field (bottom right) measurements.
The best-fit values for $g_1$ and $g_2$ give the curve indicated by the orange line.
The measurement at the lowest field is excluded from the fit.
}
\label{fig:doke}
\end{center}
\end{figure}

After S1 and S2 corrections, we determine the corrected S1 and S2 (cS1 and cS2$_b$, using only the bottom PMT for the S2) corresponding to the full absorption peak at \SI{511}{keV}.
The cS1 and cS2 are anti-correlated, as a decrease in the extracted electrons contributes to the scintillation through the recombination process.
This is described by
\begin{equation} \label{eqn:doke}
E = W \left( \frac{\rm cS1}{g_1} + \frac{{\rm cS2}_{b}}{g_2}\right),
\end{equation}
where $W = $\SI{13.7}{eV} \cite{dahl2009} and $g_1$ and $g_2$ are the photon and electron gain, respectively.
The values of $g_1$ and $g_2$ are detector-dependent and can be determined from the linearity of ${{\rm cS2}_b}/E$ as a function of ${\rm cS1}/E$.
This is shown in figure~\ref{fig:doke}.
The photopeak values of cS1 and cS2$_b$ (for example, as indicated by the red cross in figure~\ref{fig:cs1cs2}) are determined from individual  Gaussian fits to cS1 and cS2$_b$ of events in the photopeak (selected with the same procedure as outlined in section~\ref{sec:electron_lifetime}).
The cS1 uncertainty is taken as the statistical uncertainty from the Gaussian fit, while the cS2$_b$ uncertainty is dominated by the electron lifetime  uncertainty.
The measurement at \SI{100}{V} shows a significant deviation from the linear behavior and is excluded from the fit.
We attribute this to an imperfect integration of the full S2 signal due to its extreme width and low area, resulting in a low amplitude waveform that is only partially integrated by the peakfinding algorithm.
The obtained values from the fit are $g_1 = $ \num[separate-uncertainty=true]{0.102 \pm 0.003} p.e./$\gamma$ and $g_2 = $ \num[separate-uncertainty=true]{5.01 \pm 0.13} p.e./e$^{-}$, in agreement with values found in~\cite{hogenbirk2018}.

Given the values of $g_1$ and $g_2$, the S1 and S2 yields can be recalculated to absolute yields (in photons and electrons per \si{keV}, respectively), through
\begin{equation}
L_y = \frac{\left< {\rm cS1} \right> }{g_1 \cdot E}; \hspace{20mm} Q_y = \frac{\left< {\rm cS2}_b \right> }{g_2 \cdot E}.
\end{equation}

The yields depend on the incident particle type, energy and applied field.
The yield for various fields for a gamma interaction at an energy of \SI{511}{keV} is extracted from NEST and compared against our observations.
This is shown in figure~\ref{fig:superplot}(d) and \ref{fig:superplot}(e).
While the field dependence of light and charge yields is well captured by the NEST description, our measurements favor a lower light yield and higher charge yield at \SI{511}{keV} for all fields.
Note that the light and charge yield are anti-correlated as the total number of quanta $E/W$ is fixed (equation~\ref{eqn:doke}).
A best-fit description is found for a \SI{13}{\%} decrease in light yield, as shown by the dashed curves in the figures.
Changing the values of $g_1$ and $g_2$ similarly shifts the light and charge yields, however, since the uncertainty on these parameters is \SI{3}{\%}, this can only partly cause the observed discrepancy.

\subsection{Scintillation pulse shape}
In the absence of recombination luminescence, the scintillation pulse shape is described by a double exponential distribution due to the existence of two exciton states: the singlet state with lifetime $\tau_s$ and the triplet state with lifetime $\tau_t$.
Recombination luminescence has the net effect of broadening the pulse due to a delay in the formation of exciton states, and possibly a preferred formation of the triplet state for the recombination process.
This is most notable for low ionization density recoils and low electric fields \cite{kubota1979}.
Rather than constructing a detailed model of recombination, the scintillation pulse shape is usually described by an effective model, using a single exponential distribution \cite{akimov2002, dawson2005} or absorbing the delay due to recombination into an effective lifetime~$\tau_t^{\rm eff}$ \cite{xmass2016, akerib2018}.
The normalized photon emission time distribution then becomes
\begin{equation} \label{eqn:eff}
I(t; \tau_s, \tau_t^{\rm eff}, f_s) = f_s \left( \frac{1}{\tau_s} \exp \frac{-t}{\tau_s} \right) + (1 - f_s) \left( \frac{1}{\tau_t^{\rm eff}} \exp \frac{-t}{\tau_t^{\rm eff}} \right),
\end{equation}
with $f_s$ the fraction of light observed from the singlet state.

\begin{figure}[h!]
\begin{center}
\includegraphics[width=0.7\linewidth]{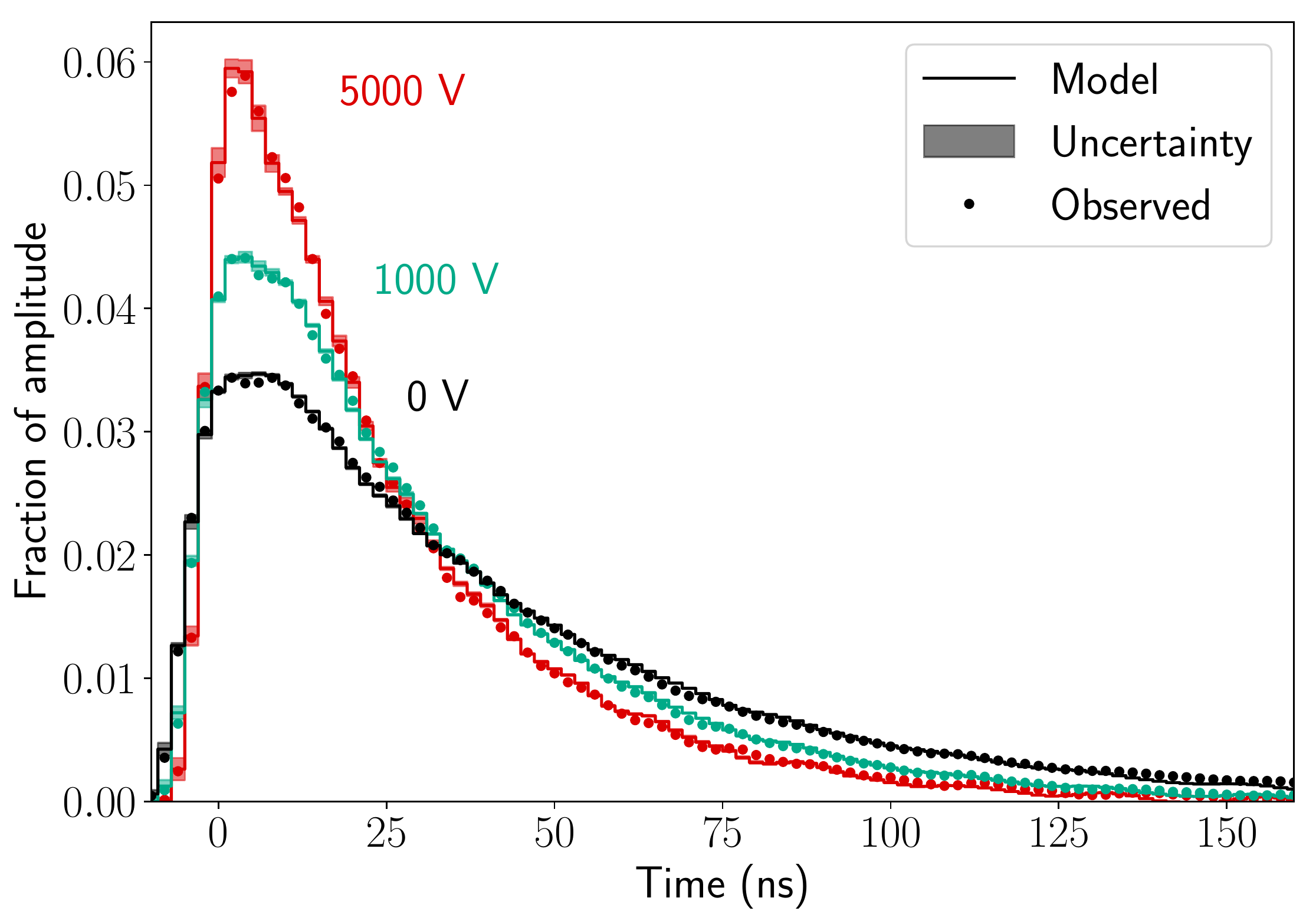}
\caption{
The average scintillation pulse shape for three of the cathode voltages.
Measured values are indicated by the points, while the solid line and the shaded area show the model description and its uncertainty, respectively.
}
\label{fig:pulsefits}
\end{center}
\end{figure}

We use the model from equation~\ref{eqn:eff} and fit this to the average pulse shape observed for all measured fields.
Only S1s from events occurring within the full absorption peak at \SI{511}{keV} and within the collimated beam position are used for this analysis.
The fitting procedure is the same as outlined in~\cite{hogenbirk2018} and uses simulated individual pulse shapes to account for the effect of pulse alignment on the average pulse shape, as well as a high time resolution single photoelectron pulse shape model.
In addition to the parameters in equation~\ref{eqn:eff}, an addition parameter $\sigma_{\rm det}$ comes from the detector time resolution, which is assumed to smear all photon detection times by a Gaussian distribution. 
This gives a total of four parameters: $\tau_s$, $\sigma_{\rm det}$ $\tau_t^{\rm eff}$ and  $f_s$.
We fix the singlet lifetime and the detector time resolution to values found in~\cite{hogenbirk2018} ($\tau_s = $ \SI[separate-uncertainty=true]{2.0 \pm 1.0}{ns} and the detector time resolution $\sigma_{\rm det} = $ \SI[separate-uncertainty=true]{1.5 \pm 0.5}{ns}).
Both are varied within their uncertainty and the effect on the best-fit values of $\tau_t^{\rm eff}$ and  $f_s$ is taken as a systematic uncertainty on these values.

Figure~\ref{fig:pulsefits} shows the average normalized waveform for three of the measured voltages.
All the waveforms are aligned such that \SI{10}{\%} of the area is at $t<0$.
The points indicate the average of the measured S1 pulse shapes.
The histograms show the best-fit model and the shaded regions around it indicate the uncertainty on the waveform that comes from the uncertainty on $\sigma_{\rm det}$ and $\tau_s$.
The curves clearly show the effect of the field on the scintillation pulse shape; much wider curves are found for low fields.

\begin{figure}[h]
\begin{center}
\includegraphics[width=0.7\linewidth]{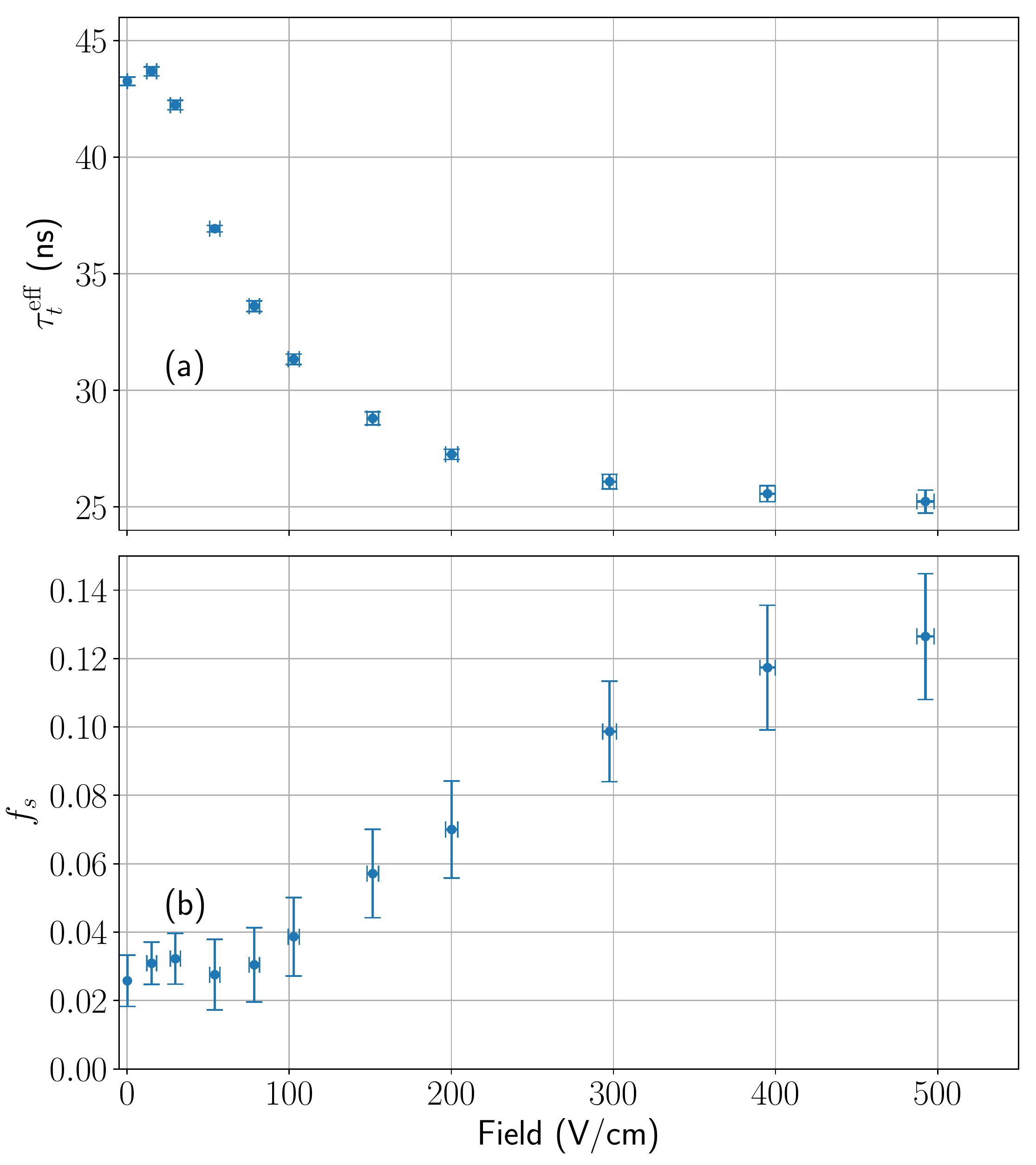}
\caption{Best-fit values of the effective triplet lifetime~$\tau_t^{\rm eff}$ (a) and the singlet fraction~$f_s$ (b) as a function of field, fit to equation~\ref{eqn:eff}.
The values of the singlet lifetime and the detector time resolution are varied within $\tau_s = $ \SI[separate-uncertainty=true]{2.0 \pm 1.0}{ns} and $\sigma_{\rm det} = $ \SI[separate-uncertainty=true]{1.5 \pm 0.5}{ns}, respectively.
The pulse shapes at low field are characterized by a long decay time of up to \SI{43.5}{ns}, and a very low singlet fraction consistent with \num{0.03} below \SI{100}{V/cm}.
At higher fields, the singlet fraction increases and the effective triplet lifetime drops to \SI{25}{ns}.
}
\label{fig:pulsefit}
\end{center}
\end{figure}

The best-fit values of $\tau_t^{\rm eff}$ and  $f_s$ as a function of field are shown in figure~\ref{fig:pulsefit}.
At low field, the pulse is characterized by a long effective triplet lifetime of approximately \SI{43.5}{ns} and only a very small contribution of the singlet fraction (\num[separate-uncertainty=true]{0.03 \pm 0.01} below \SI{100}{V/cm}), in agreement with the $\sim$\SI{45}{ns} observed in \cite{kubota1979} for $\mathscr{O}($\si{MeV}$)$ electron recoils at zero field.
We note that the lifetime at zero field is slightly below the lifetime at the lowest nonzero field.
This difference may be due to multiple scatter events that are not cut in the zero field data, as there are no S2s.
In this case, the observed scintillation light comes from multiple energy depositions, so that effectively a lower energy than \SI{511}{keV} is probed.
Since the effective triplet lifetime decreases with decreasing energy for electronic recoils, this may cause an artificially reduced effective triplet lifetime.
At high fields, the singlet fraction increases up to \num[separate-uncertainty=true]{0.13 \pm 0.02}, while the lifetime drops to \SI{25}{ns}.
The data suggest that even at high field strengths, the effective triplet lifetime is still appreciably higher than the actual triplet lifetime of \SI[separate-uncertainty=true]{22 \pm 1}{ns} measured using high ionization density tracks \cite{hitachi1983}.
This could imply that either the recombination process is still significant even at high field strengths, or that the low ionization density associated with \SI{511}{keV} gamma recoils causes a delay in the direct process of excimer formation.

\section{Conclusions}
We have measured the dependence on electric field of various quantities of interest to dual-phase liquid xenon TPCs.
For this, we used a setup with a collimated $^{22}$Na source, triggering on \SI{511}{keV} gamma recoils.
We use an electric field simulation, and whenever necessary the field inhomogeneity is taken into account for the analysis.
We measured the drift velocity, electron lifetime, diffusion constant and light and charge yields for fields ranging from \SI{10}{V/cm} up to \SI{500}{V/cm}.
We find a minor field dependence of electron lifetime, with a maximum value at \SIrange{100}{200}{V/cm}, although this may depend on the type of electron-attachment impurity.
The diffusion constant is shown to remain relatively constant for fields higher than \SI{100}{V/cm}, but rises steeply for lower fields.
The light and charge yield dependence on field is well captured by NEST, although a systematic light yield decrease of \SI{13}{\%} at \SI{511}{keV} is suggested by our data.

We fit the average scintillation pulse shape to a model containing two exponential decays, where the field-induced change of the recombination luminescence time dependence is fit by changing the effective triplet lifetime and singlet fraction.
The effective triplet lifetime reaches values up to \SI{43.5}{ns} for low fields and converges to \SI{25}{ns} at high fields, while the singlet fraction increases from \num[separate-uncertainty=true]{0.03 \pm 0.01} to \num[separate-uncertainty=true]{0.13 \pm 0.02}.

We note that the strong field dependence of the diffusion constant and the drift velocity for low fields may provide a challenge for large dual-phase TPCs if a field below \SI{100}{V/cm} is used.
The combination of a high diffusion constant and low drift velocity makes the S2s very wide, causing overlapping S2 signals for multiple scatter events if the $z$-separation for the interaction positions is not sufficiently large.
This could cause a significant increase in multiple scatter backgrounds into the single scatter signal sample, thus limiting the sensitivity to dark matter interactions.

\acknowledgments
We gratefully acknowledge support through the FOM VP139 program, financed by the Netherlands Organisation for Scientific Research (NWO).
We thank all members of the Nikhef Dark Matter group for shift duties during data-taking and for useful discussions. We are grateful for the technical support from the mechanical, electronics and computing departments at Nikhef. This work was carried out on the Dutch National e-Infrastructure with the support of the SURF Cooperative.

\bibliography{references}

\end{document}